\documentclass[14pt,a4paper,twoside,fleqn]{extarticle}
\usepackage{els-chem-crc}
\usepackage{times}

\usepackage{graphicx}
\usepackage[figuresright]{rotating}
\configuresideways
\usepackage{amssymb}
\usepackage{amsfonts}
\usepackage{amsmath}

\newcommand{\lsb} {\left[}
\newcommand{\rsb} {\right]}

\newcommand{\lab} {\left\langle}
\newcommand{\pd}{{\phantom{\dagger}}}
\newcommand{\rab} {\right\rangle}
\def\oc{\omega_{\rm{c}}}
\chapter{19}
\title{Modeling molecular conduction in DNA wires:
 Charge transfer theories and dissipative quantum transport  }

\author{R. Bulla\address[MCSD]{Theoretische Physik III, Elektronische                                                                                         Korrelationen und Magnetismus, Universit\"at Augsburg,
D-86135, Augsburg, Germany},%
         $\;$R. Gutierrez\address[reg]{Institut f\"ur Theoretische Physik, 
	Universit\"at Regensburg, D-93040 Regensburg, Germany}
        and
        G. Cuniberti\addressmark[reg]
 }

\begin{document}

\maketitle

\begin{abstract}
  Measurements of electron transfer rates as well as of
 charge transport characteristics in DNA produced a number of
 seemingly contradictory results, ranging from insulating
 behaviour to the suggestion that DNA is an efficient medium
 for charge transport. Among other factors, environmental effects 
 appear to play a crucial
 role in  determining  the effectivity of charge propagation along the 
 double helix.
 This chapter gives an overview over charge transfer theories  and 
 their  implication  for addressing  the interaction of a molecular 
 conductor with
 a dissipative environment. Further, we focus on possible applications 
 of these approaches for charge transport through DNA-based molecular 
 wires.
\end{abstract}

\section{Introduction}

The discovery of long-range electron transfer processes in 
double stranded DNA \cite{murphy93} considerably attracted  the attention of 
biologists, chemists and physicists. The motivation is threefold:
i) the possible use of  DNA-molecules in nanotechnology
applications \cite{DR}, ii) the biological role of
electron transfer in, for example, radiation damage and repair,
\cite{Boon} and iii) the potentials of biochemical sensors
based on electron transfer in DNA \cite{por04}.

Despite the intensive experimental efforts, the results for
electron {\em transport} still appear to be contradictory, ranging
from metallic conduction \cite{fink99,kas01} to insulating behaviour
with very large bandgaps \cite{porath00,storm01}. We refer the reader to 
Ref.~\cite{por04} for a recent review. The measurements of
electron {\em transfer}, on the other hand, appear to be much
better controlled and earlier discrepancies on the distance
dependence of the electron transfer rate are now attributed to the different
experimental setups \cite{Boon}.

Theoretically, several classes  of factors have been meanwhile identified, which
considerably determine the effectivity of charge propagation along the double
helix. They can be roughly classified as being related to (i) static disorder
associated with the random or quasi-random sequence of bases in DNA oligomers
\cite{roche03,roche03a,klotsa05},
(ii) dynamical disorder arising from strong structural fluctuations of the 
molecular frame \cite{hennig04,bishop02,gozema02}, and (iii) environmental effects related to the presence of 
an aqueous environment and counterions
\cite{barnett01,gervasio02,endres04,endres05,gervasio05,anantram03,anantram05}. 
While the first two factors 
can still be addressed in a first approximation by considering only 
the atomic structure of isolated 
DNA oligomers, environmental effects require the consideration of the  
solvation shells and counterions 
and their interaction with the DNA molecules. Though the
performance of {\em ab
initio} approaches has considerably improved in the last years, the description 
of the 
dynamical interaction of DNA with an environment is still a formidable
computational task
involving at least several thousands of atoms.
As a consequence, only relatively few first principle studies addressing 
this issue 
have been carried out in
the past years~\cite{barnett01,gervasio02,endres04,endres05,gervasio05,anantram03,anantram05}. 
Thus, model
Hamiltonian approaches describing charge propagation in presence of a
dissipative environment are very valuable and help to gain some insight into
the subtleties of the physical behavior  of a quantum mechanical 
system interacting 
with a macroscopic number of  degrees of freedom.

This chapter  will give an overview of different approaches to address 
charge propagation in a dissipative environment. In the next section, 
we discuss some results from {\em ab initio} calculations of DNA oligomers
in presence of an aqueous environment. In section {\bf 3} some basic facts 
on how to model the interaction between an arbitrary quantum mechanical 
system in interaction with a dissipative environment are introduced. Finally,
 in
subsection {\bf 3.1}, a special application to a DNA model is discussed.

\begin{figure}[t]
\centerline{
\includegraphics[width=4.1in,height=3.5in]{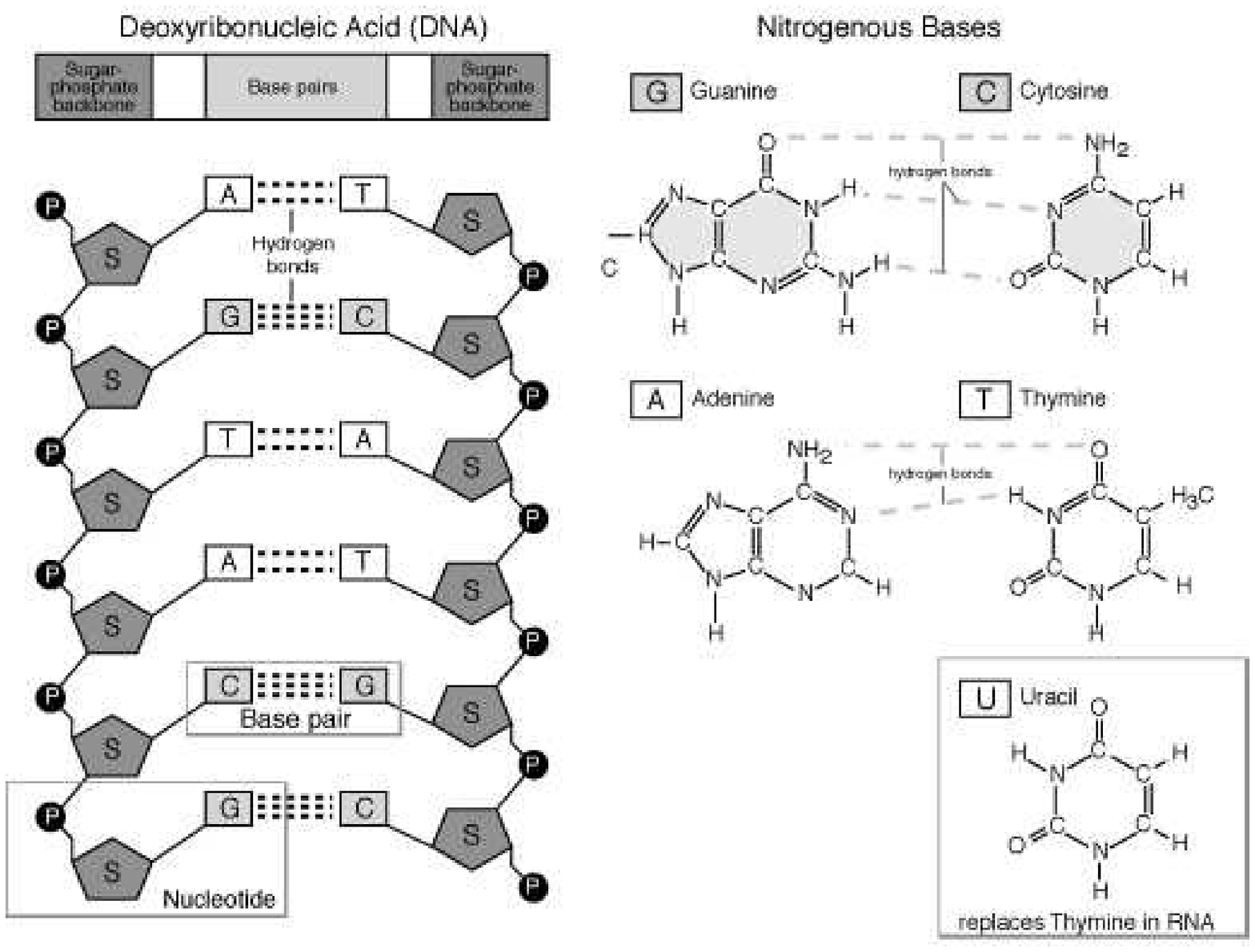}%
\includegraphics[width=1.80in,height=3.7in]{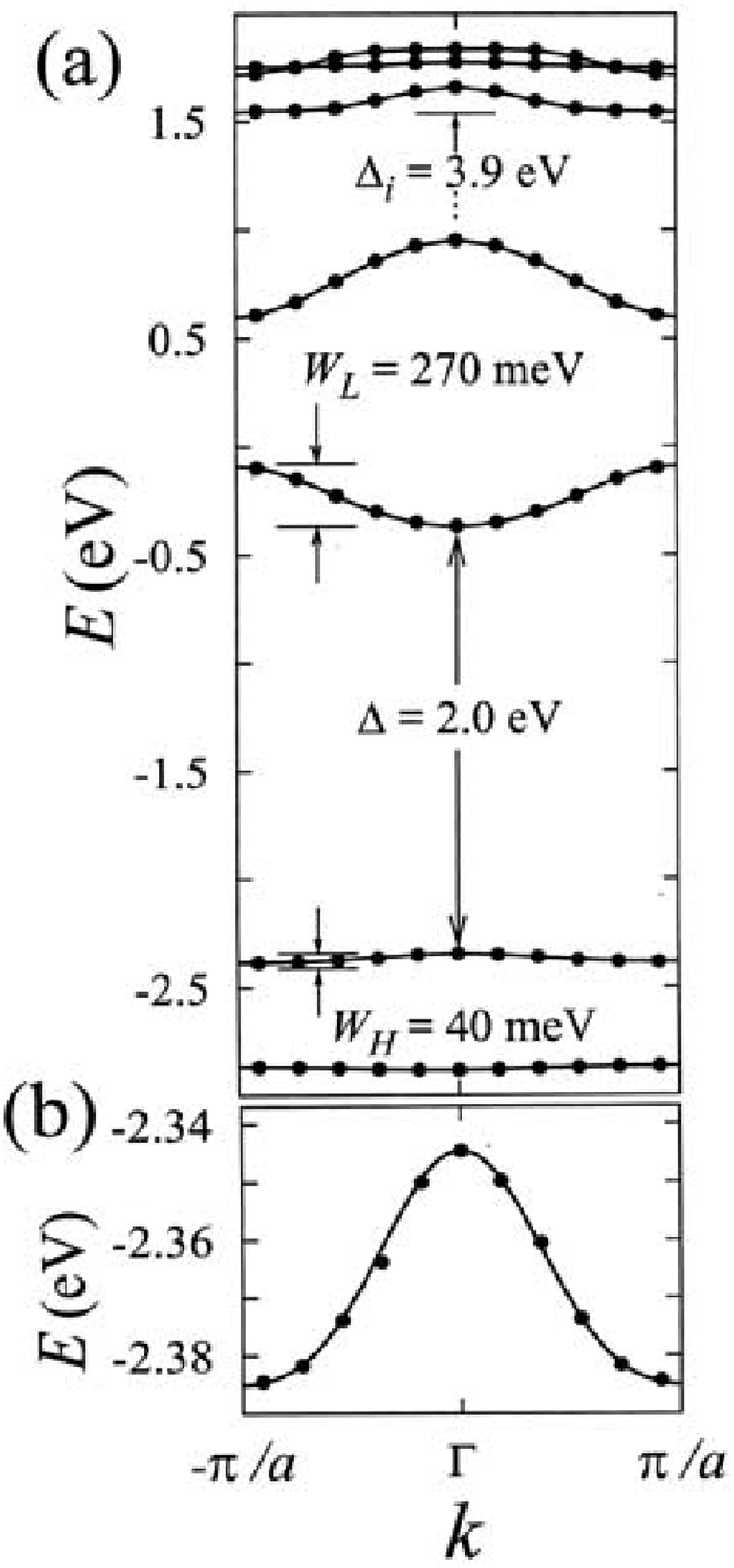}%
}
\caption{\label{fig:fig1}%
Left: Schematic representation of a double-stranded DNA oligomer with an
arbitrary base-pair sequence. 
Right: Electronic band-structure of Poly(GC) 
DNA within density-functional theory. The figures are reproduced from 
Refs.~\cite{dnapic,artacho03}  with permission.}
\end{figure}

\section{Environmental effects within {\em ab-initio} approaches}

For the purpose of illustrating some basic facts concerning the electronic 
structure of a {\em dried} DNA oligomer, let us look at a recent
band-structure calculation of poly(GC) carried out with the 
density-functional-based code
SIESTA~\cite{artacho03}. First-principle results for poly(AT) oligomers have
also 
been recently  presented~\cite{star02,star02a,star03}. The natural advantage of Poly(GC) or Poly(AT)
is its periodic structure which considerably minimizes the computational 
efforts. In Fig.~1, the resulting band-structure is shown. From the practical
point of view, it is also expected that this kind of periodic structures 
will have a higher potential applicability in molecular electronics than 
their disordered counterparts like $\lambda$-DNA.

The top-most valence band (HOMO)
and the lowest conduction band (LUMO), both having $\pi$-character, 
are separated by a bandgap of $\Delta = 2.0\,$eV.  
 The HOMO and LUMO bands are basically derived 
from the overlap of guanine and cytosine $\pi$ orbitals, respectively.
As a consequence, the charge density of  HOMO
and LUMO bands is confined along the G- and C-strands, respectively.  
The
Fermi level lies between these two bands so that the
system appears to be a insulator.

Most striking are the very small bandwidths of the bands
close to the Fermi level. The top-most valence band
has a width of $W_H = 40\,$meV while the lowest conduction band
has a somewhat broader bandwidth of $W_L = 270\,$meV.
Part (b) of Fig.~2 of Ref.~\cite{artacho03} (see Fig.~1, right panel, in this
chapter)  shows a tight-binding
modeling of the  top-most valence band using a
single orbital per base pair. The resulting 
hopping matrix elements are $t=10\,$meV (for nearest
neighbor hopping) and $t^\prime=1.5\,$meV (for next-nearest
neighbor hopping).

In principle, such a scenario allows for electronic transport
mediated by carriers which are introduced by doping
either in the top of the valence band or the bottom of
the conduction band. The extremely small bandwidths,
however, suggest that  Bloch states mediated transport cannot be stable
under the various perturbations present in DNA \cite{artacho03}.

One of the possible perturbations which have been studied
in some detail is the role of the
environment~\cite{barnett01,gervasio02,endres04,endres05,gervasio05}. 
As shown in Ref.~\cite{barnett01}, the
 existence of rather different time
scales of the environment may have a strong impact on a charge 
propagating
along the DNA molecule. First-principle  
simulations were performed, 
including four base pairs of B-DNA in the sequence GAGG, 
together with Na$^{+}$ counterions and the hydration shell. 
It turns out that holes can be gated  by the temperature dependent dynamics 
of the environment, i.~e.~there may exist 
configurations 
that localize the hole. Dynamical 
fluctuations of the
counterions  can lead, however, to configurations which support 
hole motion. A hole thus 
experiences transitions between quantum-mechanical states
 that are correlated with
different environmental configurations~\cite{remark}. These results have 
been partly  confirmed by recent {\em ab initio}
simulations in Ref.~\cite{gervasio05}. The authors have additionally 
pointed out at a
different,  proton-mediated
mechanism for hole localization, which may be quite effective in Poly(GC) DNA.

The {\em ab initio}-based studies in Refs.~\cite{gervasio02,endres04,endres05} 
 have yielded further insight 
into  the role played by 
water and counterions in modifying the low-energy electronic structure of  
DNA oligomers. Despite the differences in the DNA-conformations (Z-
\cite{gervasio02} vs. B-DNA~\cite{endres04,endres05})
 as well as in 
computational approaches (different basis sets and approximations for the
exchange-correlation potentials), they nevertheless indicate 
that the environment  can introduce midgap states. 
Though these electronic states do not form truly extended 
electronic bands, they
may support activated charge hopping at high-temperatures and thus lead to an 
enhancement of the conductivity. 
In this respect, they
resemble to some degree the defect levels induced by impurities in bulk
semiconductors.

We can conclude from this that (i) the appearance of a 
band-gap is not at all a generic feature for the band-structure
of DNA and (ii) the extremely small values of the bandwidths
do appear to be generic. The general question which arises
from that is the relation of the  bandwidths $W_H$ and
$W_L$ to other typical energy scales due to disorder effects,
electron-phonon coupling, and Coulomb-correlations. 
Furthermore, 
the environment can have a dramatic 
influence on the electronic structure of the oligomer by inducing 
defect-like states within the $\pi-\pi^{*}$ gap. However, as previously
stated, the complexity of the problem makes a full {\em ab initio} treatment
rather difficult. This leads us to the
issue of how the system-environment interaction can be modelled within a 
Hamiltonian model approach. 
Which are the essential ingredients that have to
be taken into account?

\section{Modeling the system-environment interaction}

The importance of the system-environment interaction has
long been recognized in biomolecules (such as proteins), in 
which electron transfer reactions take place. Within
Marcus theory \cite{Marcus}, the coupling of the electronic
degrees of freedom to a reaction coordinate is the first
step of a successful description of electron transfer processes.
The quantum mechanical analog of the reaction coordinate is
a phononic degree of freedom originating from vibrations of
the protein matrix. In general, there might not be one dominating
phononic mode; such a breakdown
of the standard single reaction coordinate description has been
suggested in the context of charge transfer between DNA base pairs
\cite{bruinsma00}. More importantly, even
a dominating reaction coordinate is coupled to the fluctuations
of the environment, such as surrounding water molecules, so that
the resulting spectral function $J(\omega)$ of all relevant phononic
modes can be regarded as continuous over a very broad energy range 
(in theoretical calculations, the low-energy cutoff is typically
set to $\omega=0$).

The coupling of the electronic subsystem (the electron transfered
between donor and acceptor site) to the environment
leads to a very important effect:
when an electron initially localized at the donor site tunnels
to the acceptor site (which typically has a lower energy), the
energy difference is dissipated to the environment so that the
electron transfer process is irreversible \cite{garg85}. If this friction
would be too small, or if the electron would couple to 
a single phonon mode only, the electron would oscillate
between donor and acceptor sites making the electron transfer
process highly inefficient.

In the work of Garg {\it et al.} \cite{garg85}, the friction term
has been modeled quantum mechanically via a coupling to a bath
of harmonic oscillators. A minimal model for electron transfer
processes, similar to the one proposed in \cite{garg85}, then takes the
form:
\begin{eqnarray}
  H
&=& \sum_{i={\mathrm A,D}} \varepsilon_i c^\dagger_{i} c^\pd_{i}
     - t  \left( c^\dagger_{{\mathrm D}} c^\pd_{{\mathrm A}} +
                               c^\dagger_{{\mathrm A}} c^\pd_{{\mathrm D}}
                         \right)   \nonumber \\
    &+&
      \sum_{n} \omega_{n} b_{n}^{\dagger} b^\pd_{n}
    + (g_{\mathrm A} n_{\mathrm A} + g_{\mathrm D} n_{\mathrm D}) \sum_{n}
         \frac{\lambda_n}{2} \left(  b_{n}^{\dagger} + b^\pd_{n}  \right)
    \ .
\label{eq:ebm}
\end{eqnarray}
The operators $c^{(\dagger)}_{i}$ denote annihilation (creation)
operators for electrons  on the donor ($i={\rm D}$)
and acceptor ($i={\rm A}$) sites; $n_{\rm A/D}$ is defined as
$n_i= c^{\dagger}_{i\sigma}c_{i\sigma}$.
The first two terms of the
Hamiltonian eq.~(\ref{eq:ebm}) correspond to a two-site tight-binding
Hamiltonian with $\varepsilon_i$ the on-site energies and
$t$ the hopping matrix element.

The last two terms in
eq.~(\ref{eq:ebm}) describe the free bosonic bath 
(with bosonic creation and annihilation operators $b_{n}^{\dagger}$
and $b_{n}$) and the coupling
between electrons and bosons, respectively. 
Assuming symmetric
phonon displacements due to the electronic occupancy
at donor and acceptor sites, one can set
$g_{\rm A}=1$ and $g_{\rm D}=-1$~\cite{May}.

The coupling
of the electrons to the bath degrees of freedom is completely
specified by the bath spectral function
$   J(\omega) = \pi \sum_{n}
\lambda_{n}^{2} \delta\left( \omega -\omega_{n} \right) \ $. 
The form of $J(\omega)$ can, in principle, be calculated with
molecular dynamics simulations (see, for example, \cite{Schulten2}).
To study the qualitative influence of the environment, an ohmic
bath spectral function $J(\omega)\propto\omega$ (with a suitable
high-energy cut-off) is sufficient for most cases. In this
description, dominant reaction coordinates lead to additional
resonances in the bath spectral function.

Note that such a continuous bath spectral function enforces
a {\em quantum mechanical} treatment of the phononic degrees
of freedom since the temperature range always lies within
the continuum of phononic modes.

The Hamiltonian eq.~(\ref{eq:ebm}) can be viewed as a paradigm
for modeling the system-environment interaction in biomolecules
in which the electronic degrees of freedom couple to a
dissipative environment. We should add here that the model
eq.~(\ref{eq:ebm}) can be exactly mapped onto the well-studied
spin-boson model \cite{leggett87,weiss_book} for the case of
{\em one} electron in the system. In the spin-boson model
description, the state $\vert\uparrow\rangle$ ($\vert\downarrow\rangle$)
corresponds to the electron localized at the donor (acceptor)
site. More complicated situations arise when the spin degree
of freedom of the electron --- not to be confused with the artificial spin
in the spin-boson model --- is taken into account, see the discussion
in Ref.~[\cite{bulla05}].

Calculations for these types of models in the context of
electron transfer problems have been presented in
\cite{garg85,bulla05,Egger}.  It is natural to assume
that the electron environment interaction plays an equally
important role for electron transport through the DNA double
helix. The main difference here is that electron transfer/transport
occurs over very many sites so that the two-site model eq.~(\ref{eq:ebm})
has to be suitably generalized. One such example is discussed in
the following section.

\subsection{Modeling the system-environment interaction: a DNA-wire 
in a dissipative bath }

\begin{figure}[t]
\centerline{
\includegraphics[width=4.5in,height=2.in]{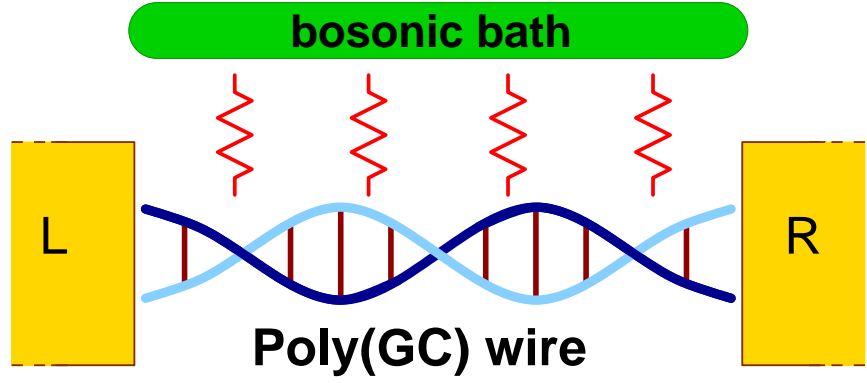}%
}
\caption{\label{fig:fig1}%
Schematic representation of a double-stranded poly(GC) oligomer coupled to left
(L) and right (R) electrodes and in interaction with a bosonic bath.}
\end{figure}

The first-principle calculations  reviewed in Section {\bf 1}  
have shown that the
environment in which DNA oligomers are placed may have a non-negligible
influence on their electronic structure. 
In this section, we will illustrate 
within an
effective model Hamiltonian approach, how the presence 
of a dissipative environment
does affect the low-energy transport properties of a DNA molecular
wire~\cite{gmc05a,gmc05b}. Our
reference system will be poly(GC) because of its periodic structure, which
should make
optimal the interbase electronic coupling along the strands. Moreover, recent 
experiments~\cite{tao04} on single poly(GC) molecules have shown non-zero current at low
bias, which is at variance with the fact that the molecule should have a (rather
large) HOMO-LUMO gap~\cite{sankey04}. Since these experiments were performed in an aqueous
environment and the authors excluded ionic current contributions, one may consider the
possibility that the environment is modifying the molecule electronic
structure.

In our model, we will exclusively focus on the low-energy transport, i.~e.~the charge
injection energies  are small compared with the molecular band gap of the
isolated molecule ($\sim 2-3 \,
eV$). Consequently, only equilibrium transport will be considered and a
transmission-like function can still be defined ~\cite{gmc05a,gmc05b,imry04}.
At low energies, only the frontier orbitals (HOMO and LUMO) of the molecule 
are expected to contribute to transport. As mentioned  in sec. {\bf 1} these
orbitals have $\pi$-character and their charge densities extend along the 
G- and C-strands for the HOMO and the LUMO, respectively.  Motivated by this, 
 we have formulated a minimal  tight-binding
 model~\cite{gmc05a,gmc05b,gio00}, where 
 a single electronic  $\pi$-orbital channel is connected to left and right 
 electrodes. The sugar-phosphate backbones are assumed to locally perturb the
 electronic states and lead to the opening of a semiconducting gap for the infinite chain. 
 As a result, the size of the  band gap can be 
 controlled by the strength of this perturbation, given by the parameter
 $t_{\perp}$ in Eq.~(2) below. The
environment is described by a collection of harmonic oscillators which linearly
couple to the charge density on the backbone sites. Assuming zero onsite
energies (the Fermi level thus lies at $E=0$), the
Hamiltonian  reads: 

\begin{eqnarray}
{\cal H} &=& -t_{\pi} \sum_{j} \lsb c^{\dagger}_{j}c_{j+1} + h.c. \rsb 
- t_{\perp}\sum_{j} \lsb b^{\dagger}_{j}c_{j} + h.c. \rsb \nonumber \\ 
&+&\sum_{\alpha} \Omega_{\alpha} B^{\dagger}_{\alpha} B_{\alpha} +\sum_{\alpha,j} 
\lambda_{\alpha} b^{\dagger}_{j}b_{j}
(B_{\alpha}+B^{\dagger}_{\alpha}) \nonumber \\ 
&+& \sum_{{\bf k}\in \rm{L,R}, \sigma} 
\epsilon_{{\bf k}\sigma} d^{\dagger}_{{\bf k}\sigma}d_{{\bf k}\sigma}  
+\sum_{{\bf k}\in \rm{L}, \sigma} ( V_{{\bf k},1} \, 
d^{\dagger}_{{\bf k}\sigma} \, c_{1} + h.c.) 
+\sum_{{\bf k}\in R, \sigma} (
V_{{\bf k},N} \, d^{\dagger}_{{\bf k}\sigma} \, c_{N} + h.c.) \nonumber \\
&=& {\cal H}_{\rm{el}} +{\cal H}_{\rm{B}}
+{\cal H}_{\rm{leads}}. \label{eq:eq1}
\end{eqnarray}

In the above equation,  
${\cal H}_{\rm{el}}={\cal H}_{\rm{c}}+{\cal H}_{\rm{b}}$ 
is the Hamiltonian of 
the HOMO (or LUMO) channel (${\cal H}_{\rm{c}}$)
and the backbone sites (${\cal H}_{\rm{b}}$), 
${\cal H}_{\rm{B}}$ contains both the 
Hamiltonian of the bath and the mutual interaction of the bath with the
electronic degrees of freedom at the backbone sites (second row). 
Finally, ${\cal H}_{\rm{leads}}$
contains the electrode Hamiltonians as well as the tunneling Hamiltonian 
describing the propagation of a charge from the leads onto the HOMO (or LUMO)
channel and viceversa. 
In absence of coupling to the bath, the eigenstates of ${\cal H}_{\rm{el}}$ yield two manifolds containing $N$ states each and separated by a
band gap, whose magnitude basically depends on the size of the tranversal 
coupling 
$t_{\perp}$.
 The bath is completely described by introducing its spectral density as 
 given by~\cite{weiss_book}:
$
J(\omega)= J_0( \frac{\omega}{\oc}) \exp^{-\omega/\oc},
\label{eq:eq8}
$
where $\oc$ is a high-frequency cut-off and we assume ohmic dissipation,
$J(\omega)\sim \omega$. By performing a unitary transformation, the linear coupling to the
bath can be eliminated. However, the transversal coupling terms will be
renormalized by exponential bosonic operators~\cite{gmc05a,gmc05b}. Using
equation of motion techniques, one can show, to lowest order in $t_{\perp}$, that
the Green function of the wire satisfies the following Dyson-equation:
\begin{eqnarray}
{\bf G}^{-1}(E)&=&E{\bf 1}-{\cal H}_{\rm{c}}-\Sigma_{\rm{L}}(E)- \Sigma_{\rm{R}}(E)
- t^2_{\perp} {\bf P}(E). \label{eq:eq6}
\end{eqnarray}

In this expression, the influence of
the electrodes  is captured by the complex self-energy functions
$\Sigma_{\rm{L/R}}(E)$. The function $P(E)$ is an entangled electron-boson 
Green function: 
$P_{j \ell}(t)=-i \Theta(t) \lab \lsb b_{j}(t){\cal
X}(t),b^{\dagger}_{\ell}(0){\cal X}^{\dagger}(0)\rsb _{+} \rab$ and
${\cal X}=
\exp{\lsb \sum_{\alpha}
(\lambda_{\alpha}/
\Omega_{\alpha})(B_{\alpha}-B^{\dagger}_{\alpha})\rsb}$. Note that $P(E)$ acts
as an additional self-energy and that the influence of the backbones is
contained only in this function. 

Several coupling regimes to the bath can be analyzed~\cite{gmc05b}. We focus
here only on the strong-coupling limit (SCL), defined by the 
condition $J_0/\oc > 1$, which basically means that the time scales of the
charge-bath interaction are much shorter compared with typical electronic time
scales. We refer the reader to
Refs.~\cite{gmc05a,gmc05b} for technical details. 

  \begin{figure}[t]
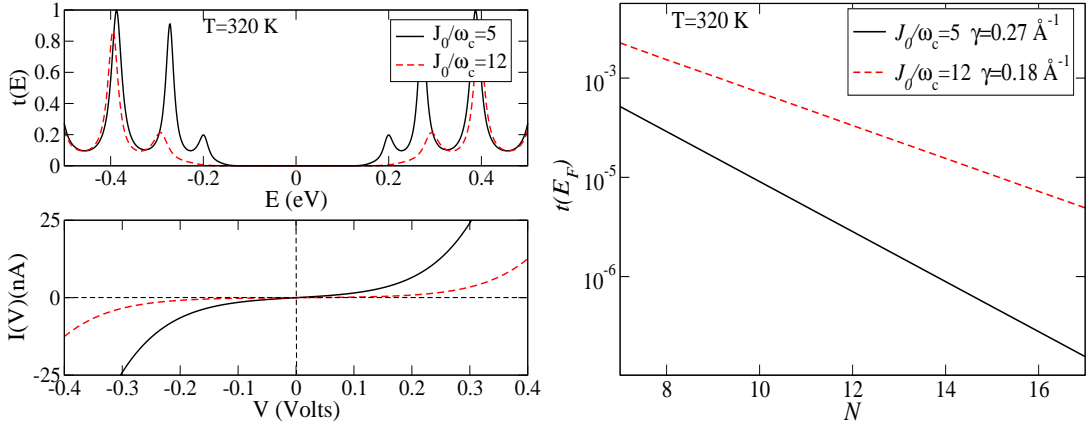

   \begin{center}
   \begin{tabular}{c}
   \includegraphics[height=2.2in,width=2.8in]{fig3a.eps}
   \includegraphics[height=2.2in,width=2.8in]{fig3b.eps}
     \end{tabular}
   \end{center}
   \caption[example] 
   { \label{fig:fig3} 
Left panel: Transmission and current for different 
electron-boson coupling strengths. At high temperatures, 
a small density of states is present at low energies. 
Right panel: correspnding dependence of the transmission at the Fermi 
energy on the number of sites $N$ in the wire. 
A very weak exponential  scaling is found, hinting at a strong contribution of 
incoherent processes. 
 }
   \end{figure} 

The impact of the bath on the electronic structure is
twofold~\cite{gmc05a,gmc05b}. On one side, the strong coupling to the bath leads
to the emergence of new bath-induced electronic states {\em inside } the wire
band gap. On the other side, however, these states are strongly damped by the
dissipative action of the bath. 
In other words, the bath  completely
destroys the coherence of transport through the wire. 
This effect has also been discussed for transport through
molecular chains under the influence of external time-dependent
fields, see \cite{Haenggi1,Haenggi2}

As a result, the bath-induced states will not manifest as resonances
in the transmission spectrum, see Fig.~3, left panel. Nevertheless, they
induce a {\em temperature dependent} background which leads to a (small) finite
density of states inside the gap. Charges injected at low energies will now 
find states supporting transport at high temperatures and thus, a finite current 
at low bias may 
flow. Hence we call the new gap a pseudogap, in
contrast to the intrinsic band gap found in the isolated wire. Note that
increasing the  interaction with the bath (increasing $J_0/\omega_{\rm c}$) does
not necessarily lead to a global increase of the current, since the frontier
orbitals of the wire are strongly damped with increasing coupling. 

Signatures of this situation are seen in the length dependence of the
transmission at the Fermi energy, see Fig. 3, right panel. Tunneling through an
intrinsic gap would lead to a very strong exponential length dependence 
$t(E_{\rm F})\sim e^{-\gamma\, N}$ with with typical inverse decay lengths  $\gamma\sim 1.5-2 \,
\rm{\AA}^{-1}$~\cite{sankey04}. We find, however, much smaller values $\sim 0.1-0.2
\, \rm{\AA}^{-1}$. With increasing bath coupling the exponential dependence
weakens, reflecting the increase of the density of states in the pseudogap and 
the strong contribution of incoherent processes~\cite{nitzan}

\section{Conclusions and Outlook}
  In conclusion, we have shown that generic model approaches can deal with experimentally relevant situations. One of the major problems for the theoretical
modeling of charge transport in DNA oligomers is the lack of a clear 
experimental picture of their transport signatures. Thus, focusing on
individual factors affecting charge propagation helps to shed light onto 
the relevant mechanisms controlling the charge dynamics in DNA. 
We have
addressed in this chapter  environmental effects in the charge transport 
through DNA oligomers within a minimal Hamiltonian model approach. 
Obviously, other factors not treated here, like 
electronic correlations, static disorder 
or internal vibrational excitations can also have a non-negligible
influence on charge propagation.

\section{Acknowledgments}
The authors thank A. Nitzan for useful comments and suggestions. 
This research was supported by the DFG
  through SFB 484, by the Volkswagen Foundation  Grant Nr. I/78-340 and by the EU under
contract IST-2001-38951.




\begin{thebibliography}{10}
\bibitem{remark} 
We note  that a similar 
situation is known in electron transfer theories, where solvent 
reorganization brings into resonance the electronic states of  donor and 
acceptor centers.
\end{thebibliography}

\end{document}